\documentstyle[aps,preprint,tighten,epsfig,amssymb]{revtex}

\begin{document}

\preprint{ \vbox{ \hbox{SNUTP-99-049} \hbox{hep-ph/9911480} }}

\draft


\title{A new anomalous trajectory in Regge theory}

\author{
Nikolai I. Kochelev$^{a}$\footnote{e-mail address: kochelev@thsun1.jinr.ru},
Dong-Pil Min$^{b}$\footnote{e-mail address: dpmin@mulli.snu.ac.kr},
Yongseok Oh$^b$\footnote{e-mail address: yoh@mulli.snu.ac.kr}, \\
Vicente Vento$^c$\footnote{e-mail address: vento@metha.ific.uv.es},
Andrey V. Vinnikov$^{d}$\footnote{e-mail address:
vinnikov@thsun1.jinr.ru}}

\address{
\bigskip
$^a$ Bogoliubov Laboratory of Theoretical Physics,
     JINR, Dubna, Moscow region, 141980 Russia\\
$^b$ Department of Physics and Center for Theoretical Physics,
     Seoul National University, Seoul 151-742, Korea\\
$^c$ Departament de F\'{\i}sica Te\`orica and Institut de F\'{\i}sica
     Corpuscular, Universitat de Val\`encia-CSIC E-46100
     Burjassot (Valencia), Spain \\
$^d$ Far Estern State University, Sukhanova 8, GSP, Vladivostok,
     690660 Russia}

\date{\today}

\maketitle
\begin{abstract}
We show that a new Regge trajectory with $\alpha_{f_1}^{} (0) \approx 1$
and slope $\alpha_{f_1}'(0) \approx 0$ explains the features of
hadron-hadron scattering and photoproduction of the $\rho$ and $\phi$
mesons at large energy and momentum transfer.
This trajectory with quantum numbers $P = C = +1$ and odd signature can
be considered as a natural partner of the Pomeron which has even
signature.
The odd signature of the new exchange leads to contributions to
the spin-dependent cross sections, which do not vanish at large energy.
The links between the anomalous properties of this trajectory, the
axial anomaly and the flavor singlet axial vector $f_1 (1285)$ meson
are discussed.
\end{abstract}



\section{Introduction}

The descriptions of the total cross sections for elastic hadron-hadron,
photon-hadron, and lepton-hadron scattering at large energy have been
important issues in the understanding of QCD.
Only a limited type of hadron reactions, the so-called hard processes,
could be understood within the perturbative QCD.
Most of the available data, however, deals with processes where the
momentum transfer between the quarks and the gluons is relatively small
and therefore their understanding should be the subject of
nonperturbative studies.
The latter are performed by the use of effective theories based on QCD.
Regge theory is a well-known nonperturbative approach to
hadron reactions \cite{collins}.
It is based on the assumption that at large energy multi-particle
exchanges with definite quantum numbers can be expressed by one effective
particle exchange with its propagator given by the so-called
Regge trajectory $(s/s_0)^{\alpha(t)}$.
It was shown that one can fix the shape of the usual Regge trajectory 
$\alpha(t) = \alpha(0) + \alpha^\prime t$ from the analysis of the mass
spectrum $M_J$ of the hadronic states with spin $J$, where
$J = \alpha(0) + \alpha^\prime M_J^2$ lies on this trajectory.
Most Regge trajectories have the same slope 
$\alpha^\prime\approx 0.9$ GeV$^{-2}$, which can be related  with the
universal value of the string tension between quarks, as has been found in 
lattice QCD calculations.
The intercepts of the usual secondary Regge trajectories are not very
large: $\alpha(0)\leq 0.5$.

The Pomeron trajectory, which was introduced into particle physics
more than forty years ago, has properties which are very different
from those of the other trajectories: It has the intercept larger
than one, $\alpha_P^{}(0)\approx 1.08$, and a slope
$\alpha_P^\prime\approx 0.25$ GeV$^{-2}$.
These peculiar properties of the Pomeron play an important role
in particle reactions, e.g., the Pomeron-exchange not only gives the main
contribution to the total hadronic cross sections but also determines the
behavior of the elastic differential cross sections at large energy and
small momentum transfer.

The origin of the Pomeron-exchange has not yet been understood from QCD,
but general wisdom ascribes its existence to the the {\it conformal
anomaly\/} of the theory.
This anomaly leads to finite hadron masses due to the non-vanishing 
of the matrix elements $\langle h| G^a_{\mu\nu} G^a_{\mu\nu} |h \rangle$
and vacuum energy density.
The Landshoff-Nachtmann model \cite{LN87} directly relates the properties
of the soft Pomeron with the non-zero value of gluon condensate
$\langle 0| G^a_{\mu\nu} G^a_{\mu\nu} |0 \rangle$  and the
nonperturbative two-gluon exchange between hadrons.

In addition to the conformal anomaly, QCD has also a non-conserved flavor
singlet axial vector current due to the {\it axial anomaly\/}.
This anomaly has played a crucial role in understanding the pseudoscalar
meson spectrum.
However only recently it has been clarified its importance for
understanding the DIS spin-dependent structure functions.
A solution to the proton spin problem (for a review see, e.g.,
Ref. \cite{AEL95}) due to the non-vanishing of the matrix
element $\langle p| G^a_{\mu\nu} \widetilde{G}^a_{\mu\nu} |p \rangle$
has been suggested, where $\widetilde{G}^a_{\mu\nu} =
\epsilon_{\mu\nu\alpha\beta} {G}^a_{\alpha\beta}/2$ is the dual
to the gluon tensor $G^a_{\mu\nu}$.
Within supersymmetric QCD only some definite chiral combinations of
these two tensors, $GG \pm i G\widetilde{G}$, correspond to
superfields \cite{KZ90}.
Therefore the matrix elements $\langle h| G^a_{\mu\nu} G^{a}_{\mu\nu}|h
\rangle$ and $\langle h| G^a_{\mu\nu} \widetilde{G}^{a}_{\mu\nu} |h
\rangle$ can be related to each other.
This link was used in Ref. \cite{KZ90} to estimate the value of the
flavor singlet axial vector charge of the proton.
Does this relation hold also in real QCD?, and if so, what are its
implications in Regge phenomenology?
If this relation holds, it would be natural that, {\it in addition to\/}
the Pomeron trajectory, a new anomalous Regge trajectory would exist,
associated to the vacuum properties of QCD through the axial anomaly
and therefore with its quantum numbers.
Our aim here is to show through data analysis that this possibility is
indeed realized.

We will estimate the contribution of this trajectory to the elastic 
proton-proton scattering cross section and show it to be very important.
We also compute the contribution of this trajectory to the photoproduction
of the $\rho^0$ and $\phi$ mesons.
Moreover we show that due to the specific quantum numbers of the possible
exchanges, polarized and unpolarized proton-proton scattering and
photoproduction of vector mesons can be very useful for investigating the
possible consequences of this new trajectory.

The remaining part of this paper is organized as follows.
In Sec.~II the connection of the properties of the $f_1$ meson with axial
anomaly is discussed.
The effect of the new trajectory in proton-proton elastic scattering
is evaluated in Sec.~III.
Section~IV is devoted to the contribution of the $f_1$-exchange to
the cross sections in $\rho^0$ and $\phi$ meson photoproduction.
In Sec.~V the contribution of the new trajectory to the polarization
observables in $pp$ collision and vector-meson photoproduction is
obtained and its nontrivial role in polarized particle reactions is
discussed.
We give a short summary and some conclusions in Sec.~VI.

\section{Properties of the $\lowercase{f}_1$ meson and the Proton Spin}

The small value of the flavor singlet axial vector charge $g_A^0$
which was measured by EMC \cite{EMC88-89}, led to the crisis
of the naive parton model for the polarized deep-inelastic scattering
\cite{AEL95}.
Despite the understanding that the fundamental origin of this phenomenon
is related to the non-conservation of the flavor singlet axial vector
current due to axial anomaly, the explicit calculation for the value of
the flavor singlet axial vector charge within QCD is absent so far
\cite{LMPR99}.
The main problem here is the poorly known nonperturbative sector of
QCD, which does not allow to perform exact calculations.
One possible way to explain the small value of $g_A^0$ is to investigate
its connection with the effects of the axial anomaly in meson
spectroscopy, and there have been many trials to connect the polarized
DIS anomaly with the anomalous properties of the $\eta^\prime$ meson,
e.g. through the U(1)$_A$ problem.
However, it should be mentioned that because of its quantum numbers
the pseudoscalar $\eta^\prime$ meson cannot give a {\it direct\/}
contribution to the double spin asymmetry in polarized DIS, which was
used to extract the value of the forward flavor singlet {\it axial vector\/}
current matrix element,
\begin{equation}
\langle p| \sum_q\bar q\gamma_5\gamma_\mu q|p \rangle = 2\, m_p g_A^0 S_\mu,
\label{me}
\end{equation}
where $m_p$ is the proton mass and $S$ is its spin.
It is evident that only flavor singlet axial vector mesons can
contribute to the left hand side of this equation.

There are three axial vector mesons with the appropriate quantum numbers
[$I^G(J^{PC}) = 0^+(1^{++})$] contribute to this matrix element: 
$f_1(1285)$, $f_1(1420)$, and $f_1(1510)$ \cite{PDG98}.
The anomalous value of the nucleon matrix element of the singlet axial
vector current implies an anomalous mixing for these axial vector mesons,
similar to what happens in the case of the pseudoscalars.
This mixing implies a strong violation of the OZI rule and therefore a
strong gluonic admixture \cite{BF96}.
{}From the point of view of the present investigation the mixing is not
important, what really matters is the role played by the lightest meson in
this channel, the $f_1(1285)$, in the whole procedure.
This meson behaves in the vector axial three isosinglet case in the same
way as the $\eta^{\prime}$ behaves in the pseudoscalar case, namely it
saturates the anomaly. 
It is the purpose of this paper to estimate the effects of the
$f_1(1285)$-exchange in elastic nucleon-nucleon scattering and 
vector-meson photoproduction, and in so doing display the special role
played by its Regge trajectory, as a consequence of the properties of
the vacuum of QCD.
Throughout this paper, $f_1$ stands for the $f_1(1285)$ state and we will
neglect the possible admixture of the other hadronic states to the wave
function of the $f_1$-meson.

Let us briefly review some of the results of Sec.~III of Ref. \cite{BF96}
to adapt them to our purpose.
The matrix element of the axial vector current can be rewritten as a sum
over all possible intermediate states by using the idea of the axial
vector dominance: 
\begin{eqnarray} \langle N| \bar q \gamma_5
\gamma_\mu q|N \rangle = \sum_{\stackrel{{\cal{A}}}{k^2\rightarrow 0}}
\frac{\langle 0| \bar q \gamma_5 \gamma_\mu q|{\cal{A}} \rangle \langle
{\cal{A}} N | N \rangle}{M_{\cal{A}}^2 - k^2},
\label{sum} 
\end{eqnarray} 
where $\cal{A}$'s are the axial vector meson states characterized
by their corresponding flavor quantum numbers.
This expression provides relations between the axial charges of the
nucleon and the coupling constants of the axial vector mesons with the
nucleon, 
\begin{equation} \langle {\cal{A}} N | N \rangle = i
g_{{\cal{A}}NN}^{} \bar u(p') \gamma_\mu \gamma_5 u(p) \varepsilon^\mu,
\end{equation} 
given by
\begin{eqnarray}
g_A^3 = \frac{\sqrt{3} f_{a_1} g_{a_1NN}^{}}{M_{a_1}^2}, \qquad
g_A^8 = \frac{\sqrt{3} f_{f_8} g_{f_8NN}^{}}{M_{f_8}^2}, \qquad
g_A^0 = \frac{\sqrt{3} f_{f_1} g_{f_1NN}^{}}{M_{f_1}^2}. 
\label{gnn} 
\end{eqnarray} 
Here the decay constants are defined by 
\begin{equation}
i\varepsilon_\mu f_A = \langle 0| \bar q
\gamma_\mu \gamma_5 q| (\bar qq)_A \rangle, 
\end{equation} 
and $\varepsilon_\mu$ is the polarization vector of the meson. 
The eighth component of the flavor octet $f_8$ will be identified with 
the $f_1(1420)$ \cite{PDG98}.

{}From the data on the decay of $\tau^- \rightarrow a_1^- + \nu_\tau$ one
obtains \cite{BF96}
\begin{equation}
f_{a_1} = (0.19 \pm 0.03) \mbox{ GeV}^2.
\label{c2}
\end{equation}
Then the first equation in Eq. (\ref{gnn}) gives
\begin{equation}
g_{a_1NN}=6.7 \pm 1.0.
\label{a1pp}
\end{equation}
The analogue with the $F$ and $D$ reduced matrix elements for the axial
vector currents allows to estimate the coupling constant of the flavor
singlet axial vector meson with the nucleon as in Ref. \cite{BF96}:
\begin{equation}
g_{f_1NN} = 2.5 \pm 0.5.
\label{gf1nn}
\end{equation}
Note if we use the SU(3) relations for the decay constants of the
axial vector meson octet, $f_{f_8}=f_{a_1}$, and the SU(6) relation
between $g_{f_1NN}^{}$ and $g_{f_8NN}^{}$, $g_{f_1NN}^{} = \sqrt{2}
g_{f_8NN}^{}$, then we obtain a larger value,
$g_{f_1NN}^{} \approx 3.5$.

The last equation in Eq. (\ref{gnn}) gives 
\begin{equation}
f_{f_1} \approx 0.11 \mbox{ GeV}^2,
\label{singlet}
\end{equation}
by using the axial vector charge
\begin{equation}
g_A^0 \approx 0.3,
\label{sidor}
\end{equation}
which was found recently \cite{LSS-SVT-FTV} by fitting the world data
on the spin-dependent structure function $g_1(x,Q^2)$.

It should be mentioned that the estimates (\ref{singlet}) and
(\ref{sidor}) are probably just {\it upper} limits for the corresponding
magnitudes, because of the uncertainties in the extrapolation of flavor
singlet part of the spin dependent structure function $g_1(x,Q^2)$
to the low $x$ region.
At any rate we find that the violation of the SU(6) symmetry for the
decay constants of axial vector mesons is very large and
$f_{a_1} \approx 2 f_{f_1}$.
A similar effect was discussed by Veneziano and Shore \cite{SV90} for
the case of pseudoscalar mesons.
They have shown that the small value of the decay constant of the flavor
singlet $\eta_0$ meson and $g_A^0$ is ascribed not to the properties of
this meson itself but to the general properties of the QCD vacuum, namely
to the phenomena of topological charge screening in the vacuum,
\begin{equation}
g_A^0 = \frac{\sqrt{6\chi^\prime(0)}}{f_\pi} g_A^8,
\label{scr}
\end{equation}
where $\chi^\prime(k^2) = d\chi(k^2)/dk^2$ and $\chi(k^2)$ is the
so-called topological susceptibility,
\begin{equation}
\chi(k^2) = i\int dx e^{ikx} \langle 0| T \frac{\alpha_s}{8\pi} G
\widetilde{G}(x) \frac{\alpha_s}{8\pi}G \widetilde{G}(0) |0 \rangle.
\label{top}
\end{equation}
Then the smallness of the decay constant $f_{f1}$ can be thought to
have the same origin, since using Eqs. (\ref{gnn}) and (\ref{scr})
one can relate this constant with $\chi^\prime(0)$,
\begin{equation}
f_{f_1} = \frac{\sqrt{3\chi^\prime(0)}}{f_\pi} f_{f_8},
\label{relation}
\end{equation}
where we have neglected the mass difference of the $f_1$ and $f_8$.

{}From all of the above discussion we conclude that the properties of 
the $f_1$ are very different from those of the other mesons that belong
to the same SU(6) multiplet, e.g. the $a_1$ and $f_1(1420)$ mesons.
Furthermore they imply a dominating behavior in the cross sections when
the $f_1$ meson contributes and which leads in general to unexpected and
thereby anomalous results.
In order to establish these facts in a concrete manner we proceed to
calculate the contribution of the $f_1$ meson for two processes, elastic
$pp$ scattering at large $|t|$ and elastic photoproduction of $\rho$ and
$\phi$ mesons.
These calculations will indeed show that the exchange of the $f_1$ meson
produces important contributions in these processes.

Before we conclude this Section we would like to compare the $f_1$ with
the anomalous Pomeron, for the reason that have been mentioned in the
introduction.
Equation (\ref{scr}) follows from the consideration of the
{\it axial anomaly} contribution to the non-conservation of the flavor
singlet axial vector current,
\begin{equation}
\partial_\mu J^{\mu 5} (x) = 2i \sum_q m_q \,\bar q \gamma_5 q
+ 2 N_f Q(x),
\label{anom}
\end{equation}
where $N_f$ is the number of flavors and
\begin{equation}
Q(x) = \frac{\alpha_s}{8\pi} G_{\mu\nu}^a \widetilde{G}^{a}_{\mu\nu}
\label{charge}
\end{equation}
is the topological charge density.
Due to its relation with the axial anomaly, the properties of the
$f_1$-exchange are related with the distribution of the topological
charge in QCD vacuum.
It is well-known that nonperturbative fluctuations of gluon fields called
instantons give rise to the nontrivial topological structure of the QCD
vacuum (see a recent review in Ref. \cite{SSh98}).
The average size of the instantons in this vacuum is much smaller than
the confinement radius,
\begin{equation}
\rho_c / R_{\rm conf} \approx 1/3.
\label{rho}
\end{equation}
Therefore the $f_1$-exchange should contribute to the hadronic cross
sections at higher values of the momentum transfer than the
Pomeron-exchange whose $t$-dependence is determined approximately
by the isoscalar electromagnetic form factor of the hadrons participating
in the process \cite{DL86}.
For the $f_1NN$ vertex, we shall use the flavor singlet axial vector form
factor
\begin{equation}
F_{f_1NN} = 1/(1-t/m_{f_1}^2)^2,
\label{aform}
\end{equation}
with $m_{f_1}=1.285$ GeV.
This is comparable to the estimate of the isosinglet axial form
factor of Ref. \cite{BKM90}, which gives $F_A^0 = 1/(1-t/1.27^2)^2$  
using the Skyrme model.
Recent experimental data \cite{LRSB99} on the flavor octet axial form
factor favor $F_A^8 = 1/(1-t/1.08^2)^2$.

This form factor decreases slower with increasing $|t|$ than the isoscalar
electromagnetic form factor of the Pomeron-nucleon vertex \cite{DL86}:
\begin{equation}
F_1(t) = \frac{4m_p^2 - 2.8t}{(4m_p^2 - t)(1-t/0.71)^2},
\label{F1t}
\end{equation}
where $1/(1-t/0.71)^2$ is the usual dipole fit.
Therefore the $f_1$ contribution will dominate over the Pomeron
contribution at large $|t|$.
In this paper we will show explicitly this behavior in elastic
proton-proton scattering and vector-meson photoproduction.

\section{Elastic hadron-hadron scattering}

One very interesting feature of the available data on hadron-hadron
scattering at high energy is their universality \cite{DL79}.
This means that at low momentum transfer $-t< 1$ GeV$^2$ and large
energy $\sqrt{s} > 20$ GeV all data are described rather well by the
Pomeron-exchange.
It was shown also that the shrinkage of the diffractive peak with
increasing energy can be explained by the following Pomeron trajectory,
\begin{equation}
\alpha_P^{}(t) = \alpha_P^{}(0) + \alpha_P^\prime t,
\label{ptraj}
\end{equation}
with the Pomeron intercept $\alpha_P^{}(0) \approx 1.08$ and slope
$\alpha_P^\prime \approx 0.25$ GeV$^{-2}$.

\begin{figure}
\mbox{} \vskip 1cm
\centering
\epsfig{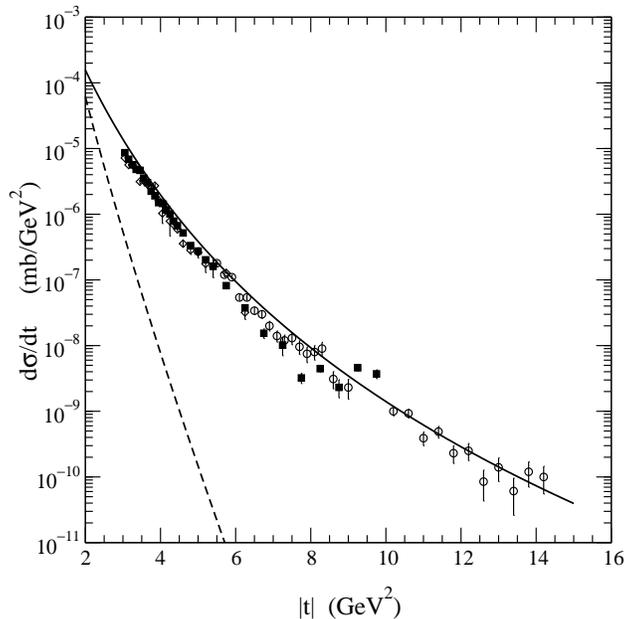}
\caption{The $f_1$-trajectory contribution to the differential cross
section for the elastic proton-proton scattering at the large energy and
momentum transfer. The solid line is the $f_1$ contribution while the
dashed line is the Pomeron contribution at $\sqrt{s} = 27.4$ GeV.
The data is from Ref.\protect\cite{CHRS78}.}
\label{fig:pp}
\end{figure}

At $-t>3$ GeV$^2$ and large energy, however, the experimental data for
the differential cross sections for elastic proton-proton and
proton--anti-proton scattering show a very different behavior.
The differential cross sections in this kinematical region do not show
{\it any\/} energy dependence and are only functions of the momentum
transfer $t$. (see Fig. \ref{fig:pp})
Several explanations for the large $|t|$ data have been suggested.
The most popular one is the Donnachie-Landshoff Odderon-exchange model
based on a perturbative three-gluon exchange mechanism between nucleon quarks
\cite{DL79,Nico99}.
In this model, however, the average momentum transfer in each quark-quark
subprocess in the experimental range of $t$ is rather small,
$-\hat{t} \approx -t/9 \leq 1.5$ GeV, and therefore justification for its
applicability is still unclear \cite{DL86}.
(See also Ref. \cite{CG82}.)
The validity of the perturbative approach to QCD to explain
nucleon-nucleon elastic scattering data is even less clear for the
Sotiropoulos-Sterman model \cite{SS94} where a six-gluon effective exchange
mechanism between nucleons was considered.

The new trajectory, which we relate to the effective $f_1$-exchange
and to the manifestation of axial anomaly in the strong interaction, 
gives a more natural explanation to this experimental data.
In this Section, we calculate the contribution of the $f_1$-exchange to
the differential cross section for proton-proton elastic scattering at
large energy, $s \gg m_p^2$ and $s \gg -t$;
\begin{equation}
\frac{d\sigma}{dt}=\frac{|T_{f_1}|^2}{16\pi s^2}.
\label{cross}
\end{equation}
The Lagrangian of the $f_1$-meson interaction with the nucleon for
a point-like vertex is
\begin{equation}
{\mathcal{L}} = g_{f_1NN}^{} \bar{\psi}_N \gamma^\mu \gamma^5
\psi_N f_\mu,
\label{lag}
\end{equation}
which allows the $f_1$-exchange to generate non-spin-flip amplitudes in 
nucleon-nucleon scattering.
The matrix element in Eq. (\ref{cross}) at $s \gg |t| \gg m_p^2$
reads
\begin{equation}
|T_{f_1}|^2 = \frac{4[g_{f_1NN}^{} F_{f_1NN}(t)]^4 s^2}{(t-m_{f_1}^2)^2},
\label{matrix}
\end{equation}
which leads to the differential cross section of the elastic
nucleon-nucleon scattering as
\begin{equation}
\frac{d\sigma}{dt} = \frac{[g_{f_1NN}^{} F_{f_1NN}(t)]^4}
{4\pi(t-m_{f_1}^2)^2}.
\label{fcross}
\end{equation}
We note that it does {\it not\/} depend on the energy $s$.

In Fig. \ref{fig:pp}, we present the contribution from the
$f_1$-exchange to the differential cross section for proton-proton
elastic scattering and compare with the experimental data at large energies,
$\sqrt{s} = 27.4$ GeV, $52.8$ GeV, and $62.1$ GeV,
and large momentum transfer $|t|>3$ GeV$^2$ \cite{CHRS78}.
In this calculation we have used $g_{f_1NN}= 2.5$, which was fixed by Eq.
(\ref{gf1nn}) through the proton spin analysis.
The result shows that the $f_1$-exchange explains the experimental
data on elastic proton-proton scattering at large momentum transfer
very well.

The Pomeron contribution to the elastic $pp$ scattering differential
cross section \cite{DL86,Land91,LM95},
\begin{equation}
\frac{d\sigma^P}{dt} = \frac{[3\beta_0 F_1(t)]^4}{4\pi} \left(
\frac{s}{s_0}
\right)^{2[\alpha_P^{}(t)-1]}
\label{pcross}
\end{equation}
with $\beta_0 \approx 2$ GeV$^{-1}$ and $s_0 = 4$ GeV$^2$ is small
at large $|t|$ compared with the $f_1$-exchange and can be neglected
in this kinematical region. 
Of course at large energy the meson-exchange should be reggeized due to
the contributions from the mesons of higher spin, $J = 3, 5, \dots$, which
belong to the same trajectory.
As discussed above, however, there is a difference in the characteristic
scales for the Pomeron and the new trajectory.
In the former, the scale is determined by the interactions between quarks
at the distance related to the confinement radius, while in the latter it
is related to the axial charge distribution (\ref{rho}).
Therefore the two trajectories should have different slopes and
\begin{equation}
\alpha_{f_1}^\prime \approx \left( \frac{\rho_c}{R_{\rm conf}} \right)^2
\alpha_P^\prime \approx 0.028 \mbox{ GeV}^{-2}.
\label{slope}
\end{equation}
The flatness of the new trajectory leads to small contributions of
the heavier mesons to the cross sections.
For example, with the slope (\ref{slope}) the next meson from the new
trajectory with $J = 3$ should have a mass $M_{J=3} \approx 9$ GeV
as determined from the equation,
\begin{equation}
J = \alpha^{}_{f_1}(0) + \alpha_{f_1}^\prime M_J^2.
\label{mass}
\end{equation}
In order to obtain a reasonable estimate we can safely neglect the
contributions from the heavier mesons and take the contribution of the
new trajectory as a fixed pole with $\alpha_{f_1}^{}(t) = 1$.

The quantum numbers of this new trajectory are determined by the quantum
numbers of the $f_1$ meson: The signature is $\sigma=-1$ with the 
parities $P = C = +1$.
Thus the new trajectory should have the same contribution to $pp$ and
$\bar pp$ collisions.
This property is different from the charge odd Odderon-exchange.
Furthermore, the strength of the interaction of the new exchange with
some hadronic state is determined by the value of the flavor singlet
axial vector charge of the hadron.
This leads to the vanishing of the contributions of this trajectory to 
the total cross sections of the elastic reactions $\pi N$, $KN$, $\pi\pi$, 
etc, because the axial vector charge of pseudoscalar mesons is zero.
One further consequence of the dominance of the new trajectory is that at 
large $|t|$ the elastic $\pi p$ cross section, that is determined by the 
Pomeron-exchange, should have very different $t$ and $s$ dependence in 
comparison with the $pp$ case.
The experimental data \cite{RBEK84} support this conclusion.%

One of the features of this new trajectory is that it is responsible
for the {\it spin dependence\/} of the lepton-hadron, photon-hadron,
and hadron-hadron cross sections at large energies.
Thus the exchange of this trajectory should determine the behavior of
the flavor singlet part of the spin-dependent structure function
$g_1^N(x)$ at $x\rightarrow 0$:
\begin{equation}
g_1^N(x) \propto \frac{1}{x^{\alpha^{}_{f_1}(0)}},
\label{g1}
\end{equation}
where $\alpha_{f_1}^{}(0) \approx 1$.
Similar behavior with 
\begin{equation}
\alpha_{f_1}^{}(0) = 0.9 \pm 0.2
\label{DIS}
\end{equation}
for the neutron structure function was found by E154 Collaboration
\cite{E154-97}.
The peculiarity of the neutron structure function $g_1^n(x)$ is the
smallness of the valence quark contribution.
Therefore the behavior of this function is determined by the flavor
singlet contribution in the experimentally available range of $x$.
For the proton target the valence quark contribution is very large and the
behavior in $x$ follows the $a_1$ trajectory contribution in this
range of $x$.
Thus the neutron target is more suitable to investigate the new
trajectory in polarized DIS.

\section{Vector Meson Photoproduction}

The recent HERA data for vector-meson photoproduction
\cite{expt,Crit99,ZEUS99} are now the subject for discussion 
in the different approaches \cite{theory}.
The interest to these data is related to the possible check of
different nonperturbative and perturbative QCD models for particle
photoproduction.
In spite of the success of several models in the description of some
properties of vector-meson photoproduction, a complete theory is
not available so far.
Indeed, the new ZEUS data for $\rho$, $\phi$, and $J/\psi$
photoproduction \cite{ZEUS99} show that the perturbative approach may
describe only $J/\psi$ photoproduction.
We also note that the nonperturbative approaches based on the dominant
contribution of the {\it universal\/} Pomeron trajectory fail to
explain the cross sections at large momentum transfer $|t|$.
Furthermore, the mechanisms which are responsible for the specific
polarization properties of the electromagnetic production of vector mesons
has not been clarified until now.
For example, it is very hard for such models to explain the large double
spin asymmetry in $\rho$ meson electroproduction at relatively small
value of $Q^2$ observed recently by the HERMES Collaboration \cite{hermes}.

\begin{figure}
\mbox{} \vskip 1cm
\centering
\epsfig{file=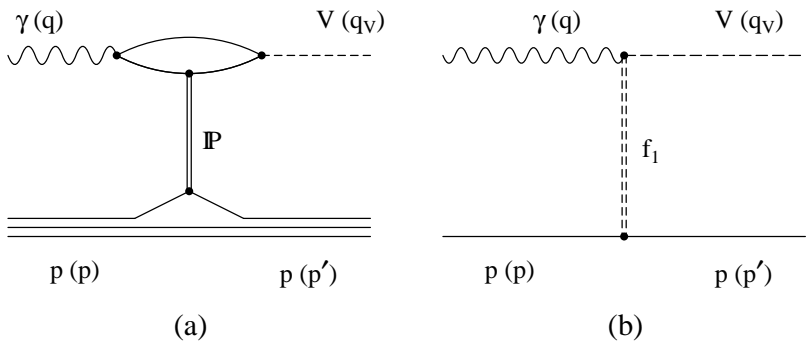, width=0.7\hsize}
\caption{Vector-meson photoproduction: (a) diffractive process by
Pomeron-exchange and (b) $f_1$-exchange.}
\label{fig:VMp}
\end{figure}

The specific features of the new trajectory can be extracted from the
experimental data of vector-meson production.
As we discussed above, because of the different $t$-dependence of the
form factors in the Pomeron-nucleon and $f_1$-nucleon vertices this
new trajectory should dominate over the Pomeron contribution at large
$|t|$ region.
To demonstrate the effect of the new trajectory we present here the
numerical results for $\rho^0$ and $\phi$ photoproduction at large
energy $W$, where $W^2 = (p+q)^2$ is the center of mass energy of the
photon-nucleon system. (Definitions of the kinematical variables can
be found in Fig.~\ref{fig:VMp}.)
We consider contributions from the two processes shown in
Fig.~\ref{fig:VMp}.
The contributions from $\pi$ and $\eta$ exchanges which give corrections
to the diffractive process at low energy \cite{TOY97} are suppressed
at large energy and are not considered in this work.

It is very well-known that the $t$-channel exchange by the Pomeron
trajectory gives the main contribution to the cross sections of
vector-meson photoproduction at small $|t|$.
The matrix element of this exchange within the Donnachie-Landshoff model
reads \cite{DL86,LM95,TOY97,DL84,PL97}
\begin{eqnarray}
T_{\lambda_V,m';\lambda_\gamma,m} &=&
i 12 \sqrt{4\pi\alpha_{\rm em}} \beta_u G_P (w^2,t)
F_1(t) \frac{m_V^2 \beta_f}{f_V} \frac{1}{m_V^2 - t}
\left( \frac{2\mu_0^2}{2 \mu_0^2 + m_V^2 - t} \right)
\nonumber \\ && \mbox{} \times
\Biggl\{ \bar{u}_{m'} (p') q\!\!\!/ \, u_m(p) \varepsilon_V^*(\lambda_V)
\cdot \varepsilon_\gamma (\lambda_\gamma)
- \left[ q \cdot \varepsilon_V^* (\lambda_V) \right]
\bar{u}_{m'} (p') \gamma_\mu u_m(p)
\varepsilon^\mu_\gamma (\lambda_\gamma)
\Biggr\},
\end{eqnarray}
where the vector-meson and photon helicities are denoted by $\lambda_V$
and $\lambda_\gamma$ while $m$ and $m'$ are the spin projections of the
initial and final nucleon, respectively,
\begin{equation}
G_P (w^2,t) = \left( \frac{w^2}{s_0} \right)^{\alpha_P^{} (t) - 1}
\exp\left\{ - \frac{i\pi}{2} [ \alpha_P^{} (t) - 1 ] \right\},
\end{equation}
with $w^2 = (2W^2 + 2 m_p^2 - m_V^2)/4$, and $F_1(t)$ is given in
Eq. (\ref{F1t}).
The vector-meson mass is $m_V$ and $\alpha_P^{} (t)$ is the Pomeron
trajectory given by Eq. (\ref{ptraj}) with
$\alpha_P^{}(0) = 1.08$ and $\alpha_P' = 1/s_0 = 0.25$ GeV$^{-2}$.
Other parameters are
\begin{equation}
\mu_0^2 = 1.1 \mbox{ GeV}^2, \qquad
\beta_u =\beta_d = 2.07 \mbox{ GeV}^{-1}, \qquad
\beta_s = 1.45 \mbox{ GeV}^{-1},
\end{equation}
and the vector-meson decay constants are $f_\rho = 5.04$ and
$f_\phi = 13.13$.

The vertex for the coupling of the axial vector current
(with momentum $q$ and polarization vector $\xi^\mu$) with two vector
currents (with momentum $k_{1,2}$ and polarization vectors
$\epsilon_{1,2}^\mu$), i.e. $1^{++}\rightarrow 1^{--}1^{--}$, can be
described by two form factors \cite{Rose63},
\begin{eqnarray}
M_{f_1VV} &=& \varepsilon_1^\alpha \varepsilon_2^{\beta *} \xi^\mu
\epsilon_{\alpha\beta\mu\delta} \left[ A_2(k_1,k_2) k_1^2 k_2^{\delta} +
A_2(k_2,k_1) k_2^2 k_1^{\delta} \right],
\label{vertex}
\end{eqnarray}
where $A_1 (k_1,k_2) = -A_2 (k_2,k_1)$.
Note that in general there are six form factors and two of them do not
contribute to the physical processes.
Here we follow the prescription of Ref. \cite{Clo98}.
Therefore, for vector-meson photoproduction by the exchange of axial
vector current, the structure of the vertex becomes very simple
in the local limit with constant $A_1(k_1,k_2)$:
\begin{equation}
V_{f_1 V \gamma} = g_{f_1 V \gamma}^{} \epsilon_{\mu\nu\alpha\beta}
\xi^\beta \varepsilon_1^\nu \varepsilon_2^\alpha k_2^2 k_1^\mu,
\label{vertex1}
\end{equation}
where $k_2^2 = m_V^2$.
This corresponds to the $AVV$ interaction Lagrangian of Ref. \cite{KM90}
obtained by using the hidden gauge approach.

The coupling constant $g_{f_1 V \gamma}^{}$ can be determined from the
experimental data on $f_1(1285) \rightarrow \gamma V$ decay \cite{PDG98}:
$\Gamma(f_1 \to \rho^0\gamma) \simeq 1.30$ MeV and
$\Gamma(f_1 \to \phi\gamma) \simeq 1.90 \times 10^{-2}$ MeV.
The vertex form of Eq. (\ref{vertex1}) gives
\begin{equation}
\Gamma(f_1 \to V \gamma) = \frac{1}{96\pi} \frac{m_V^2}{m_{f_1}^5}
(m_{f_1}^2 + m_V^2) (m_{f_1}^2 - m_V^2)^3 g_{f_1 V \gamma}^2,
\end{equation}
which leads to
\begin{equation}
|g_{f_1 \rho^0 \gamma}| = 0.94 \mbox{ GeV}^{-2}, \qquad
|g_{f_1 \phi \gamma}| = 0.18 \mbox{ GeV}^{-2}.
\end{equation}

Therefore the matrix element of the $f_1$ meson exchange to elastic
vector-meson photoproduction reads
\begin{eqnarray}
T_{\lambda_V,m';\lambda_\gamma,m} &=& i
g_{f_1 V \gamma}^{} g_{f_1NN}^{} F_{f_1NN}^{} F_{f_1 V \gamma}^{}
\frac{m_V^2}{t-m_{f_1}^2}
\epsilon_{\mu\nu\alpha\beta} q^\mu \varepsilon_V^{*\nu}(\lambda_V)
\varepsilon^\alpha_\gamma (\lambda_\gamma)
\nonumber \\ && \mbox{} \times
\left( g^{\beta\delta} - \frac{(p-p')^\beta (p-p')^\delta}{m_{f_1}^2} \right)
\bar{u}_{m'} (p') \gamma_\delta \gamma_5 u_{m} (p).
\label{coup}
\end{eqnarray}
For the form factor of the $f_1 V \gamma$ vertex, $F_{f_1 V \gamma}$,
we take
\begin{equation}
F_{f_1 V \gamma} = \frac{\Lambda_V^2 - m_{f_1}^2}{\Lambda_V^2-t},
\label{fgamma}
\end{equation}
with $\Lambda_\rho = 1.5$ GeV and $\Lambda_\phi = 1.8$ GeV.

\begin{figure}
\mbox{} \vskip 1cm
\centering
\epsfig{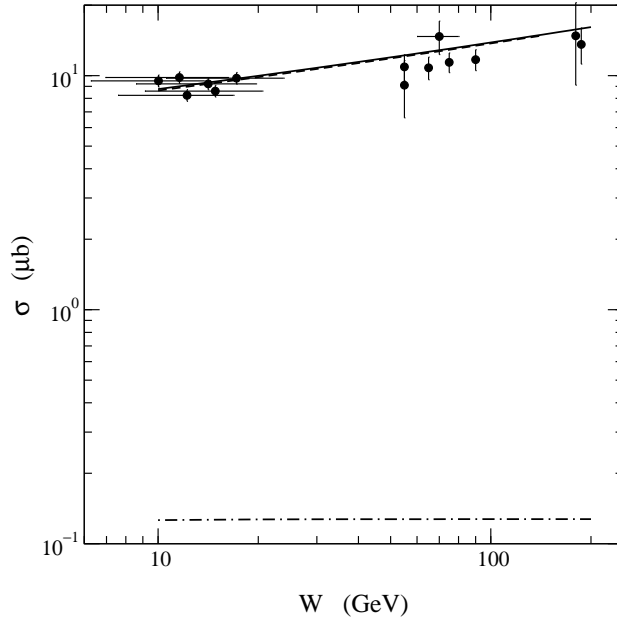}
\caption{The cross section for elastic $\rho$ meson photoproduction.
The dashed and dot-dashed lines are the contributions
from the Pomeron- and $f_1$-exchange, respectively, while the solid line
is the total cross section. The experimental data are from
Refs. \protect\cite{expt,HEPDATA}.}
\label{fig:rho}
\end{figure}

\begin{figure}
\mbox{} \vskip 1cm
\centering
\epsfig{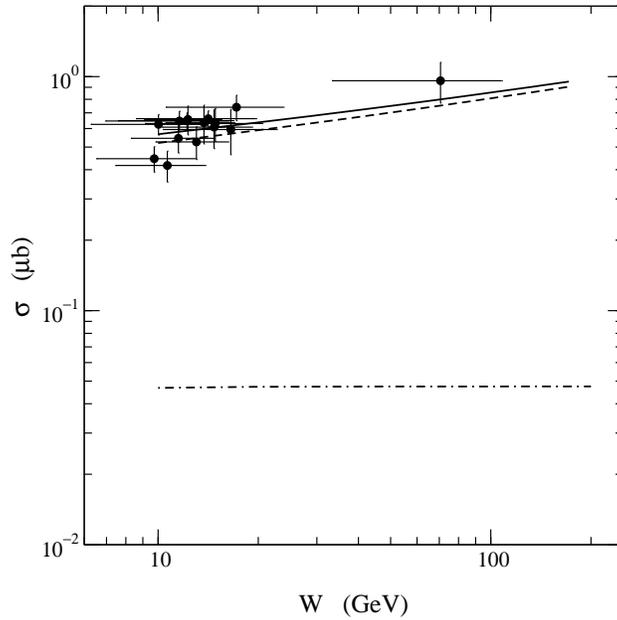}
\caption{The cross section for elastic $\phi$ meson photoproduction.
Notations are the same as in Fig. \ref{fig:rho}.}
\label{fig:phi}
\end{figure}

\begin{figure}
\mbox{} \vskip 1cm
\centering
\epsfig{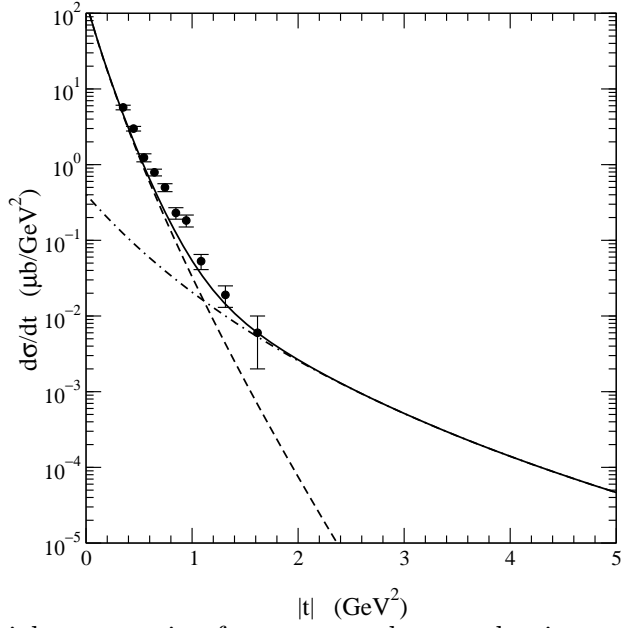}
\caption{The differential cross section for $\rho$ meson photoproduction
at $W = 94$ GeV. Notations are the same as in Fig. \ref{fig:rho}.
Experimental data are from Ref. \protect\cite{ZEUS99}.}
\label{fig:rhot}
\end{figure}

\begin{figure}
\mbox{} \vskip 1cm
\centering
\epsfig{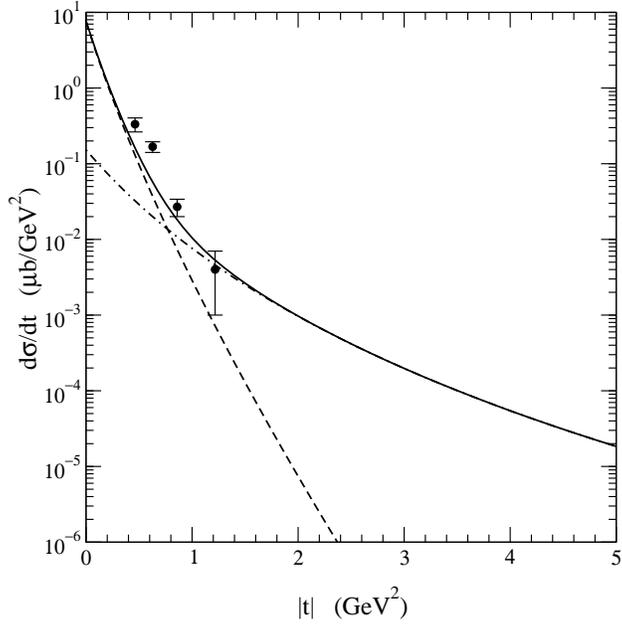}
\caption{The differential cross section for $\phi$ meson photoproduction
at $W = 94$ GeV. Notations are the same as in Fig. \ref{fig:rhot}.}
\label{fig:phit}
\end{figure}

The results for the Pomeron- and $f_1$-exchange contributions to the
total cross sections of $\rho$ and $\phi$ meson photoproduction are
presented in Figs. \ref{fig:rho} and \ref{fig:phi}.
One can find that the $f_1$-exchange contribution is nearly constant
with the energy and its magnitude is at the level of a few percent,
say $\le 5$\%, of that of the Pomeron-exchange.
Figures \ref{fig:rhot} and \ref{fig:phit} show the differential cross
sections of $\rho$ and $\phi$ meson photoproduction at $W = 94$ GeV.
These show that the contribution of the $f_1$-exchange is very large
and dominates over the Pomeron contribution in the large $|t|$ region.
Therefore, although the $f_1$ contribution is suppressed in the total
cross section, its role can be found in the differential cross sections.

\section{New Trajectory and Polarized Cross Sections}

The generic structure of the new trajectory can also be found from the
spin dependence of the related vertices, which is very different from
that of the Pomeron-exchange.
The main difference is that this exchange has unnatural parity, i.e.,
$P = (-1)^{J+1}$.
One expectation is then that the contribution of this new trajectory may
be separated by measuring spin observables, for example, the double spin
asymmetries in hadron-hadron, lepton-hadron, and photon-hadron
interactions, which vanish in the natural parity Pomeron-exchange.

The wide experimental program for investigating polarization effects in
proton-proton scattering has been suggested for RHIC \cite{RHIC}.
The main goal of this program is to extract information about polarized
parton distribution functions of the nucleon and to check the important
role of the axial anomaly in polarization physics.
We would like to emphasize that the polarized proton-proton {\it elastic}
scattering could be very useful to extract the information about
axial anomaly effects related with the contribution of the new
trajectory.

\begin{figure}
\mbox{} \vskip 1cm
\centering
\epsfig{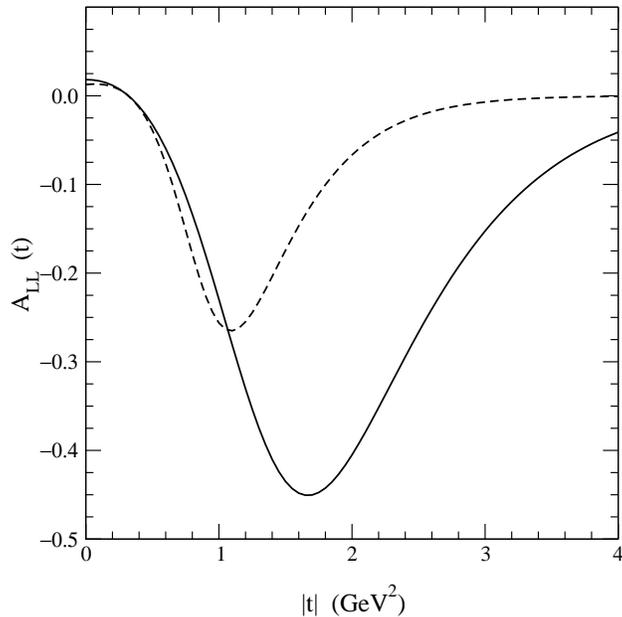}
\caption{The double longitudinal asymmetry in elastic proton-proton
scattering at $\sqrt{s} = 50$ GeV (solid line) and at $\sqrt{s} = 500$ GeV
(dashed line).}
\label{fig:aLLpp}
\end{figure}

As an example, we consider the effects of the $f_1$-exchange in the
double longitudinal spin asymmetry in the $pp$ elastic scattering:
\begin{equation}
A_{LL} =
\frac{d\sigma (\rightleftarrows) - d\sigma (\rightrightarrows)} 
{d\sigma (\rightleftarrows) + d\sigma (\rightrightarrows)} ,
\label{ALL}
\end{equation}
where $d\sigma$ denotes the differential cross section of the
proton-proton scattering and arrows show the relative orientation of
the proton spins.
This asymmetry can be written through the helicity amplitudes
\cite{KS76},
\begin{eqnarray}
&& \Phi_1 = <++|++>, \qquad \Phi_2 = <++|-->, \qquad
\Phi_3 = <+-|+->, \nonumber \\
&& \Phi_4 = <+-|-+>, \qquad \Phi_5 = <++|+->,
\end{eqnarray}
as
\begin{equation}
A_{LL} = \frac{-|\Phi_1|^2-|\Phi_2|^2+|\Phi_3|^2+|\Phi_4|^2}  
{|\Phi_1|^2+|\Phi_2|^2+|\Phi_3|^2+|\Phi_4|^2+4|\Phi_5|^2}.
\end{equation}
It is very well-known that at large energy $s \gg |t|$ one can neglect
the contribution to the cross sections from the spin-flip amplitudes,
$\Phi_2$, $\Phi_4$, and $\Phi_5$.
Furthermore, the Pomeron- and $f_1$-exchanges have different relations
for the $\Phi_1$ and $\Phi_3$ amplitudes:
\begin{equation}
\Phi_1^P=\Phi_3^{P}, \qquad \Phi_1^{f_1}=-\Phi_3^{f_1}.
\label{con}
\end{equation}
Therefore only the interference between the two exchanges can lead to
non-vanishing $A_{LL}$
\begin{equation}
A_{LL} \approx -\frac{2\Phi_1^{f_1} \, \mbox{Re}(\Phi_1^P)}
{|\Phi_1^P|^2+|\Phi_1^{f_1}|^2},
\label{as3}
\end{equation}
where we have neglected the contribution from the imaginary part of the
$f_1$-exchange to $A_{LL}$ because of the very small deviation of the
intercept of new trajectory from one.
Without this interference the asymmetry is suppressed as $t/s$.
The final result for the asymmetry can then be written simply as
\begin{equation}
A_{LL} = \frac{2\sqrt{d\sigma^Pd\sigma^{f_1}}}{d\sigma^P+d\sigma^{f_1}}
\, \sin\left\{ \frac{\pi}{2}[\alpha_P^{} (t)-1] \right\},
\label{as4}
\end{equation} 
where $d\sigma^{P(f_1)}$ is the contribution from the Pomeron- and
$f_1$-exchange, respectively, to the differential cross section.
The result of the calculation for the two RHIC energies,
$\sqrt{s} = 50$ GeV and $500$ GeV is presented in 
Fig.~\ref{fig:aLLpp}.
This shows the large asymmetry at $|t| \le 4$ GeV$^2$ and the weak
dependence on the energy, which is related to the high intercept
of the new trajectory.

Another very useful quantity which is also sensitive to the new
trajectory is the difference of the total cross sections of polarized
initial states:
\begin{equation}
\Delta\sigma_{L} = \sigma (\rightleftarrows) -
\sigma (\rightrightarrows).
\label{sLL}
\end{equation}
This asymmetry can be written in the terms of the helicity amplitudes,
\begin{equation}
\Delta\sigma_{L} = \frac{1}{2\sqrt{s(s-4m^2)}}\, \mbox{Im} \left[
\Phi_1(0) - \Phi_3(0) \right],
\label{total1}
\end{equation}
using the optical theorem.
It should be mentioned that only unnatural parity exchange can contribute
to $\Delta\sigma_{L}$.
Since the Pomeron-exchange has $\Phi_1=\Phi_3$, it does not contribute to
$\Delta\sigma_{L}$ although it dominates in the unpolarized total cross
section.
Thus at large energy only the new anomalous trajectory which has unnatural
parity can contribute.
Since the asymmetry $\Delta\sigma_{L}$ is proportional to the imaginary
part of the new trajectory, it vanishes if the intercept of the $f_1$
trajectory is $\alpha_{f_1}(0) = 1$ exactly.
Therefore the measurement of this asymmetry at large energy gives the
opportunity to measure the deviation of $\alpha_{f_1}(0)$ from one:
\begin{equation}   
\Delta\sigma_{L}=\frac{2g^2_{f_1NN}}{m_{f_1}^2}
\,\sin\left\{ \frac{\pi}{2} [\alpha_{f_1}^{} (0)-1] \right\}.
\label{sec3}
\end{equation}

Recently the E581/704 Collaboration \cite{E704} has reported a measurement
on $\Delta\sigma_{L}$ for proton-proton scattering at rather large momentum
$p = 200$ GeV$/c$:
\begin{equation}   
\Delta\sigma_{L} = -42 \pm 48 \pm 53  \ \mu b.
\label{exp}
\end{equation}
{}From this result, with assumption that at this energy the contribution of
other Regge trajectories with unnatural parity is small, we can estimate
the intercept of the new trajectory as
\begin{equation}
\alpha_{f_1}^{}(0) = 0.99 \pm 0.04,
\label{int2}
\end{equation}
which is in good agreement with Eq. (\ref{DIS}), which was extracted from
the analysis on the behavior of the structure function $g_1^n(x)$ in
polarized deep-inelastic scattering at low $x$.

\begin{figure}
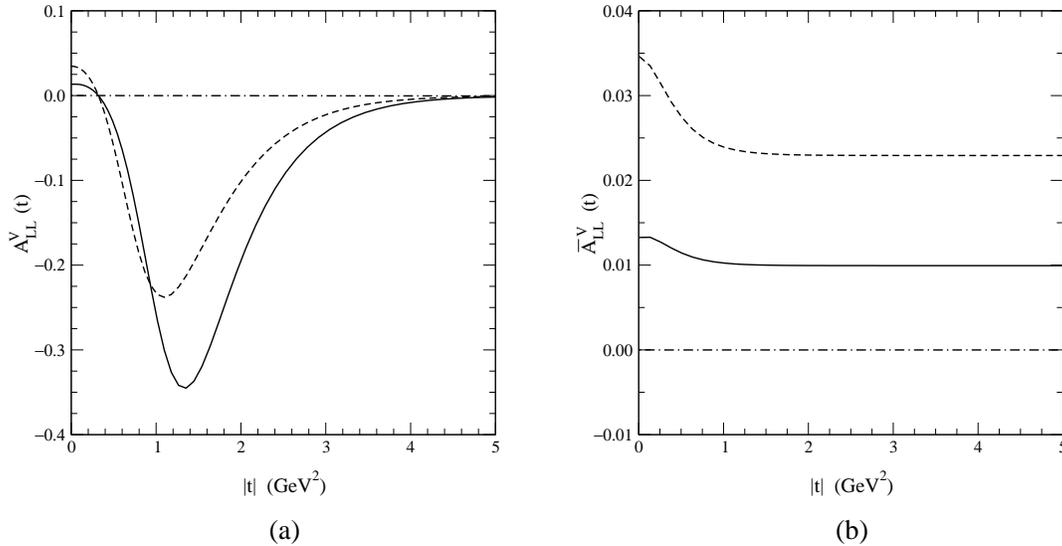

\mbox{} \vskip 1cm
\centering
\epsfig{file=fig8a.eps, width=0.4\hsize} \qquad
\epsfig{file=fig8b.eps, width=0.4\hsize}
\caption{The beam-target double asymmetry of vector-meson
photoproduction: (a) $A^V_{LL} (t)$ and (b) $\bar{A}^V_{LL}(t)$
at $W = 100$ GeV.
The solid and dashed lines correspond to the results with the
$f_1$-exchange in $\rho$ and $\phi$ production, respectively.
The dot-dashed lines are the results of the Pomeron-exchange model for
both cases.}
\label{fig:polbt}
\end{figure}

Another suitable field to look for the effects of this trajectory is the
polarized vector-meson photoproduction. 
Given in Fig. \ref{fig:polbt}(a) are the results of the double spin
asymmetries in $\rho$ and $\phi$ photoproduction for longitudinally
polarized proton targets and polarized photon beams at $W = 100$ GeV.
We define the beam-target double asymmetry as
\begin{equation}
A_{LL}^V (t) = \frac{d\sigma(\rightleftarrows) -
d\sigma(\rightrightarrows)}
{d\sigma(\rightleftarrows) + d\sigma(\rightrightarrows)},
\label{asym}
\end{equation}
where the arrows denote relative orientations of the proton spin and photon
helicity.
In Fig. \ref{fig:polbt}(b) we also give $\bar{A}_{LL}^V(t)$ defined as
\begin{equation}
\bar{A}_{LL}^V (t) = \frac{\tilde\sigma(\rightleftarrows) -
\tilde\sigma(\rightrightarrows)}
{\tilde\sigma(\rightleftarrows) + \tilde\sigma(\rightrightarrows)},
\end{equation}
where
\begin{equation}
\tilde{\sigma}(t) \equiv \int_{|t|_{\rm min}}^{|t|} dt'
\frac{d\sigma}{dt'}.
\end{equation}
One can find that the values of $A_{LL}^V$ and $\bar{A}_{LL}^V$ are very
different from those of the Pomeron-exchange model.
The $f_1$-exchange contribution gives rise to nonvanishing values
for $A_{LL}^V$ at small $|t|$ region, while the Pomeron model predicts
$A_{LL} \approx 0$ at this region.
It should be mentioned that recently the first measurements
of the double spin asymmetries for vector-meson electroproduction
was reported by the HERMES Collaboration \cite{hermes}.
They found large asymmetry for $\rho$ meson electroproduction with $|t|
\le 0.4$ GeV$^2$.
The asymmetries of $\phi$ and $J/\psi$ electroproduction could not be
measured accurately and cannot give any definite conclusion to
these asymmetries.
Nevertheless, the data on the double spin asymmetry of $\rho$
electroproduction could not be explained by the Pomeron-exchange model
and we propose the $f_1$-exchange as a possible candidate which
gives nonvanishing double spin asymmetries at high energy.

\section{Summary and Conclusions}

We have shown that the $f_1 (1285)$ meson plays a very special role in
some hadronic processes in the large energy and large momentum transfer
region due to its special relation to the axial anomaly through the
matrix elements of the axial vector current. 
This behavior is reminiscent of another effective particle with a very
dominant behavior in hadron physics in this region, the Pomeron, whose
relation with the conformal anomaly seems to be the motivation behind
this behavior.
Therefore it seems natural to describe the properties of the lightest
$f_1$ in terms of Regge theory by associating to this resonance an
anomalous trajectory, the odd signature companion of the even signature
Pomeron.
Our analysis of proton-proton scattering and vector-meson photoproduction
has confirmed our suspicions.
The contribution of the $f_1$-exchange to the cross sections of these
processes does not depend on the energy, a clear signature of its anomalous
Regge behavior.

We have found that the $f_1$-meson exchange can explain the differential
cross sections of proton-proton scattering at large energy and large
momentum transfer.
The difference in the spin structure of the vertices of the
$f_1$-trajectory from the Pomeron-exchange leads to their different
contributions to the polarized proton-proton scattering and polarized
vector-meson photoproduction.
As a result, this trajectory gives non-zero values for the beam-target
double spin asymmetries in both processes, and can be distinguished from
the Pomeron-exchange model.
These examples show the importance of the new trajectory in hadron
reactions, and can be tested, for example, by the $pp2pp$ experiment 
at RHIC \cite{Guryn}, where a wide program to measure 
various spin-dependent elastic and total cross sections 
has been suggested. (See recent discussions in Ref. \cite{LT99}.)

As another test of the new trajectory, we suggest to investigate $b_1
(1235)$ photoproduction, which decays mostly into $\omega\pi$.
Since the $b_1$ has quantum numbers $I^G(J^{PC}) = 1^+ (1^{+-})$,
the empirical Gribov-Morrison rule \cite{Morr67} prohibits the
Pomeron-exchange in this process, and because of the $C$ parity
vector-meson-exchanges such as $\omega$ and $\rho$ cannot contribute.
The possible pseudoscalar-meson exchange contribution, however, gives
usually decreasing total cross section with the initial energy, while
experiments \cite{Omega84} show approximately constant total cross
section at large energies.
Therefore, we expect that the new trajectory related with the $f_1$ may
give important contribution to $b_1$ photoproduction.
But the currently available data on this reaction \cite{Omega84} are very
limited and new experiments at current electron facilities are strongly
called for.

We also suggest to analyze the asymmetry $P_\sigma$ of the two-meson
decays of vector-mesons produced by linearly polarized photon beams
\cite{SSW70}.
Since this asymmetry is $+1$ for natural parity exchange and $-1$
for unnatural parity exchange, the $f_1$ contribution would be
found from deviations of this quantity from $+1$ at large energy.

In summary, in the energy region of the analyzed experiments two 
trajectories, the Pomeron and the $f_1$, dominate the total cross
sections and the asymmetries.
And, therefore, this new anomalous Regge trajectory, which is responsible
for spin effects at large energy, should be considered as a natural odd
signature companion to the Pomeron.

\acknowledgements

We are grateful to A.~E. Dorokhov, S.~B. Gerasimov, E.~A. Kuraev,
E.~Leader, and T.~L. Trueman for illuminating discussions.
Y.O. is grateful to M. Tytgat for useful informations.
D.-P.M. and Y.O. were supported in part by the KOSEF through the CTP of
Seoul National University.
V.V. was supported by DGICYT-PB97-1227, ERB FMRX-CT96-008, and 
the Theory Division at CERN, where part of this work was done.


\end{document}